\documentclass[11pt,a4paper]{article}
\usepackage{amsmath}
\usepackage{amssymb}
\usepackage{amscd}
\usepackage[dvips]{graphicx}

\numberwithin{equation}{section}

\begin{document}

\newtheorem{thm}{Theorem}[section]
\newtheorem{cor}{Corollary}[section]
\newtheorem{lem}{Lemma}[section]
\newtheorem{prop}{Proposition}[section]
\newtheorem{defn}{Definition}[section]
\newtheorem{exas}{Examples}[section]
\newtheorem{exam}{Example}[section]
\newtheorem{counterexam}{Counterexample}[section]
\newtheorem{rem}{Remark}[section]
\newtheorem{ques}{Question}[section]
\newtheorem{conj}{Conjecture}[section]
\numberwithin{equation}{section}


\title{Extended inequalities for weighted Renyi entropy involving generalized Gaussian densities}

\author{Salimeh Yasaei Sekeh}
\date{}
\maketitle

\begin{abstract}
In this paper the author analyzes the weighted Renyi entropy in order to derive several inequalities in weighted case. Furthermore, using the proposed notions $\alpha$-th generalized deviation and $(\alpha,p)$-th weighted Fisher information, extended versions of the moment-entropy, Fisher information and Cram\'{e}r-Rao inequalities in terms of generalized Gaussian densities are given.
\end{abstract}


\def\wt{\widetilde}
\def\fB{\mathfrak B}\def\fM{\mathfrak M}\def\fX{\mathfrak X}
 \def\cB{\mathcal B}\def\cM{\mathcal M}\def\cX{\mathcal X}
\def\be{\mathbf e}
\def\bu{\mathbf u}\def\bv{\mathbf v}\def\bx{\mathbf x} \def\by{\mathbf y} \def\bz{\mathbf z}
\def\om{\omega} \def\Om{\Omega}
\def\bbP{\mathbb P} \def\hw{h^{\rm w}} \def\hwi{{h^{\rm w}}}
\def\beq{\begin{eqnarray}} \def\eeq{\end{eqnarray}}
\def\beqq{\begin{eqnarray*}} \def\eeqq{\end{eqnarray*}}
\def\rd{{\rm d}} \def\Dwphi{{D^{\rm w}_\phi}}
\def\BX{\mathbf{X}}\def\Lam{\Lambda}\def\BY{\mathbf{Y}}
\def\BZ{\mathbf{Z}} \def\BN{\mathbf{N}}

\def\tM{\tilde M}
\def\mwe{{D^{\rm w}_\phi}}
\def\DwPhi{{D^{\rm w}_\Phi}} \def\iw{i^{\rm w}_{\phi}}
\def\bE{\mathbb{E}}
\def\1{{\mathbf 1}} \def\fB{{\mathfrak B}}  \def\fM{{\mathfrak M}}
\def\diy{\displaystyle} \def\bbE{{\mathbb E}} \def\bu{\mathbf u}
\def\BC{{\mathbf C}} \def\lam{\lambda} \def\bbB{{\mathbb B}}
\def\bbR{{\mathbb R}}\def\bbS{{\mathbb S}}
 \def\bmu{{\mbox{\boldmath${\mu}$}}}
 \def\bPhi{{\mbox{\boldmath${\Phi}$}}}  \def\bPi{{\mbox{\boldmath{$\Pi$}}}}
  \def\btheta{{\mbox{\boldmath${\theta}$}}}
 \def\bbZ{{\mathbb Z}} \def\fF{\mathfrak F}\def\bt{\mathbf t}\def\B1{\mathbf 1}
\def\hwphi{h^{\rm w}_{\phi}}
\def\BW{\mathbf{W}} \def\bw{\mathbf{w}}
\def\beal{\begin{array}{l}}
\def\beac{\begin{array}{c}}
\def\beacl{\begin{array}{cl}}
\def\ena{\end{array}}
\def\WBJ{\mathbf{J}^{\rm w}_{\phi}}
\def\BS{\mathbf{S}}
\def\BK{\mathbf{K}}
\def\BB{\mathbf{B}}
\def\wtD{{\widetilde D}}

\def\mwe{{D^{\rm w}_\phi}}
\def\DwPhi{{D^{\rm w}_\Phi}} \def\iw{i^{\rm w}_{\phi}}
\def\bE{\mathbb{E}}
\def\1{{\mathbf 1}} \def\fB{{\mathfrak B}}  \def\fM{{\mathfrak M}}
\def\diy{\displaystyle} \def\bbE{{\mathbb E}} \def\bu{\mathbf u}
\def\BC{{\mathbf C}} \def\lam{\lambda}
\def\bbB{{\mathbb B}} \def\bbM{{\mathbb M}}
\def\bbR{{\mathbb R}}\def\bbS{{\mathbb S}}
\def\blam{{\mbox{\boldmath${\lambda}$}}}
\def\bmu{{\mbox{\boldmath${\mu}$}}} \def\bta{{\mbox{\boldmath${\eta}$}}}
\def\bzeta{{\mbox{\boldmath${\zeta}$}}}
 \def\bPhi{{\mbox{\boldmath${\Phi}$}}}  \def\bPi{{\mbox{\boldmath{$\Pi$}}}}
 \def\bbZ{{\mathbb Z}} \def\fF{\mathfrak F}\def\bt{\mathbf t}\def\B1{\mathbf 1}
\def\hwphi{h^{\rm w}_{\phi}}
\def\BT{{\mathbf T}} \def\BW{\mathbf{W}} \def\bw{\mathbf{w}}
\def\beal{\begin{array}{l}}
\def\beac{\begin{array}{c}}
\def\beacl{\begin{array}{cl}}
\def\ena{\end{array}}
\def\WBJ{\mathbf{J}^{\rm w}_{\phi}}
\def\BS{\mathbf{S}}
\def\BK{\mathbf{K}}
\def\tL{\mathbf{L}}
\def\BB{\mathbf{B}}
\def\vphi{{\varphi}}
\def\rw{{\rm w}}
\def\bZ{\mathbf Z}
\def\wtf{{\widetilde f}} \def\wtg{{\widetilde g}} \def\wtG{{\widetilde G}}
\def\vphi{\varphi}
\def\rT{{\rm T}}
\def\tA{{\tt A}} \def\tB{{\tt B}} \def\tC{{\tt C}} \def\tI{{\tt I}} \def\tJ{{\tt J}} \def\tK{{\tt K}}
\def\tL{{\tt L}} \def\tP{{\tt P}} \def\tQ{{\tt Q}} \def\tS{{\tt S}}
\def\fB{\mathfrak B}\def\fM{\mathfrak M}\def\fX{\mathfrak X}
 \def\cB{\mathcal B}\def\cM{\mathcal M}\def\cX{\mathcal X}
\def\bu{\mathbf u}\def\bv{\mathbf v}\def\bx{\mathbf x}\def\by{\mathbf y}
\def\om{\omega} \def\Om{\Omega}
\def\bbP{\mathbb P} \def\hw{{h^{\rm w}}} \def\hwphi{{h^{\rm w}_\phi}}
\def\beq{\begin{eqnarray}} \def\eeq{\end{eqnarray}}
\def\beqq{\begin{eqnarray*}} \def\eeqq{\end{eqnarray*}}
\def\rd{{\rm d}} \def\Dwphi{{D^{\rm w}_\phi}}
\def\BX{\mathbf{X}}
\def\hwphiii{{h^{\rm w}_{\phi_1\otimes\phi_2\otimes\dots\otimes \phi_n}}} \def\hwphii{{h^{\rm w}_{\phi_1\otimes\phi_2}}}
\def\mwe{{D^{\rm w}_\phi}}
\def\DwPhi{{D^{\rm w}_\Phi}} \def\iw{i^{\rm w}_{\phi}}
\def\bE{\mathbb{E}}
\def\1{{\mathbf 1}} \def\fB{{\mathfrak B}}  \def\fM{{\mathfrak M}}
\def\diy{\displaystyle} \def\bbE{{\mathbb E}}
\def\fB{\mathfrak B}\def\fM{\mathfrak M}\def\fX{\mathfrak X}
 \def\cB{\mathcal B}\def\cM{\mathcal M}\def\cX{\mathcal X}
\def\bu{\mathbf u}\def\bv{\mathbf v}\def\bx{\mathbf x}\def\by{\mathbf y}
\def\om{\omega} \def\Om{\Omega}
\def\bbP{\mathbb P} \def\hw{h^{\rm w}} \def\hwi{{h^{\rm w}}}
\def\beq{\begin{eqnarray}} \def\eeq{\end{eqnarray}}
\def\beqq{\begin{eqnarray*}} \def\eeqq{\end{eqnarray*}}
\def\rd{{\rm d}} \def\Dwphi{{D^{\rm w}_\phi}}
\def\BX{\mathbf{X}}\def\Lam{\Lambda}\def\BY{\mathbf{Y}}

\def\mwe{{D^{\rm w}_\phi}}
\def\DwPhi{{D^{\rm w}_\Phi}} \def\iw{i^{\rm w}_{\phi}}
\def\bE{\mathbb{E}}
\def\1{{\mathbf 1}} \def\fB{{\mathfrak B}}  \def\fM{{\mathfrak M}}
\def\diy{\displaystyle} \def\bbE{{\mathbb E}} \def\bu{\mathbf u}
\def\BC{{\mathbf C}} \def\lam{\lambda} \def\bbB{{\mathbb B}}
\def\bbR{{\mathbb R}}\def\bbS{{\mathbb S}} \def\bmu{{\mbox{\boldmath${\mu}$}}}
 \def\bPhi{{\mbox{\boldmath${\Phi}$}}}
 \def\bbZ{{\mathbb Z}} \def\fF{\mathfrak F}\def\bt{\mathbf t}\def\B1{\mathbf 1}
\def\UX{\underline{X}}
\def\ux{\underline{x}}
\def\Hw{H^{\rm w}}
\def\OY{\overline{Y}} \def\oy{\overline{y}}
\def\OY{\overline{Y}} \def\oy{\overline{y}} \def\omu{\overline{\mu}} \def\OSigma{\overline{\Sigma}}
\def\bbS{\mathbb{S}}
\let\phi\varphi
\section{The weighted $p$-Renyi entropy}

$\quad$In 1960, Renyi was looking for the most general definition of information measures that would preserve the additivity for independent events and was compatible with the axioms of probability. He started with Cauchy's functional equation and he ended up with  the general theory of means. The consequence of this investigation derived the definition of Renyi entropy, see \cite{R}. In addition, Stam \cite{S} showed that a continuous random variable with given Fisher information and minimal Shannon entropy must be Gaussian. However the moment-entropy inequality established the result saying: a continuous random variable with given second moment and maximal Shannon entropy must be Gaussian as well. Cf. \cite{F, Z, LYZ1, CT}.

Furthermore, The Cram\'{e}r-Rao inequality shows that the second moment of a continuous random variable is bounded by the reciprocal of its Fisher information. Cf. \cite{Cr}. Currently in \cite{LYZ}, the notions of relative Renyi entropy, ($\alpha$,$p$)-th Fisher information, as a general form of Fisher information associated with Renyi entropy including $\alpha$-th moment and deviation has been introduced. More results and certain applications emerged in \cite{CHV, FU, Be1, Be2}. Later, an interesting generalization of Stam's inequality involving the Renyi entropy was given, check again \cite{LYZ}, wherein it has been asserts that the generalized Gaussian densities maximizes the Renyi entropy with given generalized Fisher information.

The initial concept of the weighted entropy as another generalization of entropy was proposed in \cite{BG, G, C}. Certain applications of the weighted entropy has been presented in information theory and computer science (see \cite{ShMM, SiB, SuYS, SuStKel, SuYSSt}).


Let us now give the definition of the weighted entropy. For given function $x\in\bbR\mapsto\phi (x)\geq 0$, and an RV
$X:\;\Om\to\bbR$, with a PM/DF $f$,
the weighted entropy (WE) of $X$ (or $f$) with  weight function (WF) $\phi$ is defined by
\beq\label{eq:1.1}\hwphi (X)=\hwphi (f) =-\int_\bbR\phi (x )f(x )\log\,f(x)\rd x=-\bbE_X(\phi\log\,f)\eeq
whenever the integral $\diy\int_\bbR\phi (x )f(x )\Big(1\vee|\log\,f(x)|\Big)\rd x<\infty$. (A
standard agreement $0=0\cdot\log\,0=0\cdot\log\,\infty$ is adopted throughout the paper). Next, for two functions, $x\in\bbR\mapsto f(x)\geq 0$ and $x\in\bbR\mapsto g(x)\geq 0$,
the relative WE of $g$ relative to $f$ with WF $\phi$ is defined by
\beq\label{eq:1.2}
\Dwphi (f\|g)=\int_\bbR\phi (x )f(x )\log\frac{f(x )}{g(x )}\rd x.
\eeq
When $\phi\equiv 1$ the relative WE yields the Kullback-leibler divergence. we refer the reader \cite{BG, C, SY}. Following the same argument as in the WE, here, we propose a generalization form of the WE named {\it $p$-th weighted Renyi entropy} (WRE). As we said, the aim of this work is to extend the aspects introduced in \cite{LYZ} to the weighted case. On the other hand, the famous assertions like the moment-entropy, the Fisher information and the Cram\'{e}r-Rao inequalities are generalized to the weighted case referring the generalized Gaussian densities. One sees if $\phi\equiv 1$ the results coincide with the standard forms.\\
In author's opinion, this work is devoted in a similar way as analysing the Renyi entropy by elaborating newly established assertions for the WE. This leads several interesting bounds stem from the particular cases of $\alpha$ and $p$. Let us begin with the $p$-th WRE's definition.
\begin{defn}\label{Def:1.1}
The $p$-th weighted Renyi entropy (WRE) of a RV $X$ with probability density function (PDF) $f$ in $\bbR$, given WF $\phi$ and for $p>0,p\neq 1$, is defined by
\beq h^{\rm w}_{\phi,p}(X):=h^{\rm w}_{\phi,p}(f)=\diy\frac{1}{1-p}\log \int_{\bbR} \phi(x) f^p(x)\rd x.\eeq
Observe that if $\varphi\equiv 1$, the WRE, $h^{\rm w}_{\phi,p}(f)$, becomes the known Renyi entropy, denoted by $h_p(f)$, cf. \cite{R}. Moreover, we propose the $p$-th weighted Renyi entropy power (WRP) of a PDF $f$ by
\beq\label{WREP} N^{\rm w}_{\phi,p}(f)=\exp\big(h^{\rm w}_{\phi,p}(f)\big).\eeq
Here and below we assume that integrals are with respect to Lebesgue measure over the real line $\bbR$ and  absolutely convergent. Throughout the paper we also suppose that the entropies are finite as well as the expectation $\bbE_f[\phi]$. Observe that
\beq\label{Eq:1.5} \lim\limits_{p\rightarrow 1}N^{\rm w}_{\phi,p}(f)=N^{\rm w}_{\phi,1}(f)=\exp\bigg(\diy \frac{h^{\rm w}_{\phi}(f)}{\bbE_f[\phi]}\bigg).\eeq
On the other words as ${p\rightarrow 1}$ the WRP intends to the weighted entropy power (WEP), see \cite{CT, KS2, SYK}.
Note that both $h^{\rm w}_{\phi,p}(f)$ and $N^{\rm w}_{\phi,p}(f)$ are continuous functions in $p$.\\
\begin{rem}
Recalling the definition of $(\alpha,p)$-th Fisher information (FI) of PDF $f$ with derivative function $f'$:
\beq\label{Def:FI} \Big(J_{\alpha,p}(f)\Big)^{\beta p}=\diy \int_\bbR |f^{p-2}f'|^\beta f\;\rd x,\quad \hbox{for}\quad \beta\in (1,\infty], \alpha\in(1,\infty),\eeq
and $\alpha^{-1}+\beta^{-1}=1$. Particularly $\varphi=|f'|^\beta\big/f$ reads an obvious relation
\beqq N^{\rm w}_{\phi,p}(f)=\Big(J_{\alpha,r}(f)\Big)^{\beta r/(1-p)}\quad \hbox{where}\quad r=(p+2\beta-2)/p.\eeqq
\end{rem}
This explains a practical fact that all drived assertions in terms of $p$-th WRE can be applied for the $(\alpha,p)$-th FI as well. For more details refer \cite{Be1, Be2, FU}.\\
\begin{rem} Given PDF $f$ and the WF $\varphi$, set $\chi=\bbE_f[\varphi]$. For weighted PDF $f_\varphi:=\varphi\;f\big/\chi$, one yields
\beqq h^{\rm w}_{\varphi^p,p}(f)=\diy h_{p}(f_\varphi)+\diy\frac{p}{1-p}\log \chi.\eeqq
\end{rem}
Next, extending the standard notions, we can also introduce the {\bf relative} $p$-th WRP of $f$ and $g$: for $p>0,p\neq 1$ and given WF $x\in\bbR\mapsto\phi(x)\geq 0$
\beq\label{RPWRP} N^{\rm w}_{\phi,p}(f,g)=\diy \frac{\bigg(\diy\int_{\bbR} \phi\; g^{p-1}f \rd x\bigg)^{1/(1-p)}\bigg(\diy\int_{\bbR}\phi\; g^p \rd x\bigg)^{1/p}}{\bigg(\diy\int_{\bbR}\phi\; f^p\rd x\bigg)^{1\big/p(1-p)}}.\eeq
Further, more generally, for given two functions $f$ and $g$ we employ the relative $p$-th WRP and define the relative $p$-th WRE of $f$ and $g$ by
\beq D^{\rm w}_{\phi,p}(f\|g)=\log N^{\rm w}_{\phi,p}(f,g).\eeq
Then one can obtain
\beq D^{\rm w}_{\phi,p}(f\|g)=\diy\frac{1}{1-p}\log\Big(\int_{\bbR}\varphi\;g^{p-1}\;f\;\rd x\Big)+\frac{1-p}{p} h^{\rm w}_{\varphi,p}(g)-\frac{1}{p} h^{\rm w}_{\varphi,p}(f). \eeq
Therefore when $p\rightarrow 1$, one yields
\beq\label{Eq:1.1} N^{\rm w}_{\phi,1}(f,g)=\lim\limits_{p\rightarrow 1}N^{\rm w}_{\phi,p}(f,g)=\exp\Bigg\{\diy \frac{\Dwphi (f\|g)}{\bbE_f[\phi]}\Bigg\}.\eeq
 \end{defn}

Evidentally as $p\rightarrow 1$, the relative $p$-th WRE does not imply the relative WE accurately, although it is a proportion of $\Dwphi (f\|g)$, (\ref{Eq:1.1}). Going back to (\ref{RPWRP}), the relative $p$-th WRE is continuous in $p$.
\vskip .5 truecm
\begin{thm}
Given non-negative PDFs $f$, $g$ and WF $\phi$, one has\\
{\rm (a)} If $p>0$, $p\neq 1$
 \beq\label{nonnegativity:RWRE}\quad D^{\rm w}_{\phi,p}(f\|g)\geq 0.\eeq
{\rm (b)} For $p=1$, given WF $\phi$, suppose the following inequality
\beq\label{Eq:assume1} \bbE_g[\phi]\leq \;\bbE_f[\phi]\eeq
is fulfilled, then (\ref{nonnegativity:RWRE}) holds true, that is $D^{\rm w}_{\phi,1}(f\|g)\geq 0$. In both (a), (b), equality occurs iff $f\equiv g$.
\end{thm}
\vskip .5 truecm
{\bf{Proof}} \; The case $p=1$ follows by using the Gibbs inequality, cf \cite{SY}.
\beqq \begin{array}{l}
- D^{\rm w}_{\phi,1}(f\|g)\\
\quad \leq \diy\frac{1}{\bbE_f[\phi]}\int_{\bbR} \phi\;f\;\1(f>0)\;\bigg[\frac{g}{f}-1\bigg]\rd x\\
\quad \leq \diy\frac{1}{\bbE_f[\phi]}\bigg\{\int_{\bbR}\phi\;g\;\rd x-\int_{\bbR}\phi\;f\;\rd x\bigg\}\leq 0,\quad \hbox{owing to (\ref{Eq:assume1})}.\end{array}\eeqq
If $p>1$, owing to the H\"{o}lder inequality (see Lemma 1 cf. \cite{LYZ}), one can write
\beqq \begin{array}{ccl}\diy \int_{\bbR} \phi\; g^{p-1} f\;\rd x&=&\diy \int_{\bbR} \Big(\phi^{\frac{1}{p}} g \Big)^{p-1}\;\Big(\phi^{\frac{1}{p}} f\Big)\;\rd x\\
&\leq& \diy\bigg(\diy \int_{\bbR} \phi\;g^{p}\rd x\bigg)^{\frac{p-1}{p}}\;\bigg(\diy \int_{\bbR}\phi\; f^p\;\rd x\bigg)^{\frac{1}{p}}.\end{array}\eeqq
Next for $p<1$ one also derives from the H\"{o}lder inequality:
\beqq \begin{array}{ccl}
\diy \int_{\bbR}\phi\; f^p\;\rd x&=&\diy \int_{\bbR}\Big(\phi\;g^{p-1}\;f\Big)^p\;\Big(\phi\;g^p\Big)^{1-p}\;\rd x\\
&\leq& \diy \bigg(\diy \int_{\bbR}\phi\; g^{p-1}\;f\;\rd x\bigg)^{p}\; \bigg(\diy \int_{\bbR}\phi\;g^p\;\rd x\bigg)^{1-p}.\end{array}\eeqq
The equality takes place from the equality in the H\"{o}lder inequality. $\quad$ $\blacksquare$

\begin{exam}
Let the WF be $\phi(x)=\exp(\gamma x)$, $\gamma\in\bbR$. Consider the PDFs  $f\sim {\rm Exp}(\lambda_1)$, $g\sim {\rm Exp}(\lambda_2)$ such that $\lambda_1,\lambda_2>0$. Then for $p>0$, $p\neq 1$  under the following list of suppositions:
\beqq \lambda_2(p-1)+\lambda_1-\gamma>0 \quad\hbox{and}\quad \lambda_i\;p-\gamma>0,\quad i=1,2.\eeqq
we compute
\beq\label{Eq:exam1}\begin{array}{ccl} D^{\rm w}_{\phi,p}(f\|g)&=& \diy \frac{1}{1-p}\log \Big(\frac{\lambda_1\lambda_2^{p-1}}{\lambda_2(p-1)+\lambda_1-\gamma}\Big)\\
&&-\diy \frac{1}{p}\log\Big(\frac{\lambda_2^p}{\lambda_2p-\gamma}\Big)-\diy\frac{1}{p(1-p)}\log\Big(\frac{\lambda_1^p}{\lambda_1p-\gamma}\Big).\end{array}\eeq
Pictorially it can be seen that for $p\neq 1$, (\ref{Eq:exam1}) takes non-negative values. Now for $\gamma<0$ implement (\ref{Eq:assume1}):
\beq\label{Eq:exam2} \lambda_1(\lambda_2-\gamma)\geq \lambda_2(\lambda_1-\gamma)\Rightarrow \lambda_1\geq \lambda_2\eeq
In particular choose $p=1$, applying some straightforward calculations leads
\beq D^{\rm w}_{\phi,1}(f\|g)=\diy \log\frac{\lambda_1}{\lambda_2}+\diy\frac{\lambda_2-\lambda_1}{\lambda_1-\gamma}.\eeq
Here also $\gamma<0$. Performing simple numerical simulation, one can show the behavior of $D^{\rm w}_{\phi,1}(f\|g)$ for chosen value $\lambda_1$, $\lambda_2$ over selected range of $\gamma$. It can be indicated that for instance if $\gamma\in(-5,-1)$, in special case $\lambda_1=0.1$, $\lambda_2=1$, the both inequalities (\ref{Eq:assume1}) and (\ref{nonnegativity:RWRE}) in $p=1$ are violated. Whereas one can demonstrate when $\gamma\in(-0.04,-0.01)$, $\lambda_1=0.1$ and $\lambda_2=0.2$, the assertion (\ref{nonnegativity:RWRE}) holds true while the assumption (\ref{Eq:assume1}) is not satisfied. Further, with the same forms of $f$ and $g$, we observe that bounds (\ref{nonnegativity:RWRE}) and (\ref{Eq:assume1}) are fulfilled with $\lambda_1=3.5$, $\lambda_2=1.5$ and $\gamma\in(-10,-1)$.
\end{exam}
\begin{defn}\label{Def:1.2}
Given the WF $x\in \bbR\mapsto\phi(x)\geq 0$ and $\alpha\in (0,\infty)$ we introduce the $\alpha$-th {\bf generalized moment} of a PDF $f$ as follows:
\beq \mu_{\phi,\alpha}(f)=\diy \int_{\bbR}\phi(x)|x|^\alpha f(x)\; \rd x.\eeq
Considering that the integral exists. Sequently define the $\alpha$-th {\bf generalized deviation} of the PDF $f$ for $\alpha\in [0,\infty]$:
\beq\label{Def:sigma} \sigma_{\phi,\alpha}(f)=\left\{\begin{array}{ll}
  \exp\bigg(\diy \frac{1}{\bbE_f[\phi]}\int_{\bbR}\phi(x) f(x) \log |x|\;\rd x\bigg) & \alpha=0, \\
  \big(\mu_{\phi,\alpha}(f)\big)^{1/\alpha} & \alpha\in(0,\infty), \\
  esssup \big\{\phi(x)|x|: f(x)>0\big\} & \alpha=\infty.
\end{array}\right.\eeq
Note that all definitions are valid when the integrals exist, on the other hand the expressions are finite. Also it is worthwhile to mention that $\sigma_{\phi,\alpha}(f)$ is continuous in $\alpha$. Observe that taking $\phi\equiv 1$, (\ref{Def:sigma}) coincides with the standard $p$-th deviation, (6) in \cite{LYZ}.
\end{defn}
\begin{rem}
Assume for $\alpha\in(0,\infty)$ the function $g(x)=f(x)|x|^\alpha\big/\chi:\bbR\mapsto \mathbb{C}$, where $\chi:=\bbE_f[|x|^\alpha]$, is integrable. For any real number $\xi$, let $\varphi(x)=\diy e^{-2\pi i x\xi}$. Then $\alpha$-th generalized moment of $f$ becomes a proportion of the Fourier transform for $g$, i.e. $\widehat{g}(\xi)$. Also in special case $\varphi(x)=\diy e^{itx},\; t\in\bbR$ the $\mu_{\phi,\alpha}(f)$ equals the proportion of characteristic function for $g$.\\

Furthermore, consider $\varphi$ addresses to the non-negative polynomial function of $|x|$, that is for which constants $a_0,\dots,a_n$ that
\beq\label{Varphi:Poly} \varphi(x)=a_n|x|^n+a_{n-1}|x|^{n-1}+\dots+a_1|x|+a_0.\eeq
and $\phi\geq 0$. Then one obtains
\beqq \mu_{\phi,\alpha}(f)=\diy \sum_{i=0}^n a_i\; \mu_{\alpha+i}(f).\eeqq
Here $\mu_{\alpha+i}(f)$ stands $\alpha+i$-th moment of $f$ defined in \cite{LYZ}.
\end{rem}
\begin{exam}
Let $\varphi(x)$ takes the form (\ref{Varphi:Poly}), $X\sim Exp(\lambda)$. Then
$$\sigma_{\varphi,\alpha}(f)=\diy\bigg(\sum_{i=0}^n a_i \diy\frac{(\alpha+i)!}{\lambda^{\alpha+i}}\bigg)^{1/\alpha}.$$
In spite of $\alpha$-th moment, this function is not always increasing in $\alpha\in(0,\infty)$. For instance let $n=3$, $a_0=1,\; a_1=-2,\; a_2=-1,\;a_3=2$ and $\lambda\in(0.5,1.2)$, then $\sigma_{\varphi,\alpha}(f)$ decreases in the range $\alpha\in(1,2)$.
\end{exam}

\section{Generalized $p$-Gaussian densities}

Let us continue the paper with weighted information measures of generalized $p$-Gaussian densities. For this, let $X$ be a RV in $\bbR$. Then for $\alpha\in[0,\infty]$ and $p>1-\alpha$, the generalized $p$-Gaussian has the PDF
\beq\label{PDF:G} G(x)=\diy\left\{\begin{array}{ll}
a_{\alpha,p}\big(1+(1-p)|x|^\alpha\big)^{1/(p-1)}_+ &\quad p\neq 1,\\
a_{\alpha,1}\exp\big\{-|x|^\alpha\big\} & \quad p=1,
\end{array}\right.\eeq
here we use the notation $\big(x\big)_+=\max\{x,0\}$ and
\beqq a_{\alpha,p}=\diy \left\{\begin{array}{ll}
\diy \frac{\alpha(1-p)^{1/\alpha}}{2 \beta(\frac{1}{\alpha},\frac{1}{1-p}-\frac{1}{\alpha})}  &\quad p<1,\\
\diy \frac{\alpha}{2\Gamma(\frac{1}{\alpha})} & \quad p=1,\\
\diy \frac{\alpha(p-1)^{1/\alpha}}{2 \beta(\frac{1}{\alpha},\frac{p}{p-1})} & \quad p>1.
\end{array}\right.\eeqq
Note that for $p>1$ and $\alpha =0$ the $G$ is defined almost every $x\in \bbR$ as
\beq G(x)=a_{0,p}\big(-\log|x|\big)^{1/(p-1)}_+,\qquad\hbox{here}\quad a_{0,p}=\diy 1\big/2\Gamma(\frac{p}{p-1}). \eeq
In addition, in case $\alpha=\infty$ and $p>0$, the PDF $G$ is given as follows:
\beq G(x)=\left\{\begin{array}{ll}
\diy \frac{1}{2} &\quad |x|\leq 1,\\
0 & \quad |x|>1, \qquad \quad \hbox{here}\;\; a_{\infty, p}=\frac{1}{2}. \end{array}\right. \eeq

In the context of communication transmission, a model PDF can characterize the statistical behaviour of a signal. For multimedia signals, the generalized Gaussian distribution is often used, cf. \cite{Oh}. The generalized Gaussian can be applied to model the distribution of Discrete Cosine Transformed (DCT) coefficient, the wavelet transform coefficient, pixel coefficient and so on. Thus, it might be used in video and geometry compression. This distribution is also known in economic as Generalized Error Distribution (GED). We emphasize that the generalized Gaussians are also the first-dimensional version of the extremal functions for sharp Sobolev, log-Sobolev and Gagliardo-Nirenberg inequalities. Cf. \cite{DD}.\\

Now, passing to the generalized weighted Fisher information we establish the following definition.
\begin{defn}\label{Def:2.1}
For given $\alpha\in[1,\infty]$ and $p\in \bbR$, the $(\alpha,p)$-th {\bf weighted Fisher information} (WFI) of a PDF $f$ denoted by $J^{\rm w, \phi}_{\alpha,p}(f)$ is defined in different cases: If $\alpha\in (1,\infty)$, let $\beta\in (1,\infty]$ be the H\"{o}lder conjugate of $\alpha$, $\alpha^{-1}+\beta^{-1}=1$. The $J^{\rm w,\phi}_{\alpha,p}(f)$ is proposed by
\beq\label{def.GWFI} J^{\rm w,\phi}_{\alpha,p}(f)=\diy \int_{\bbR} \phi\;|f^{p-2}f'|^\beta\; f\; \rd x. \eeq
Note that the PDF $f$ is absolutely continuous. If $\alpha=1$ then $\Big(J^{\rm w,\phi}_{\alpha,p}(f)\Big)^{1/\beta}$ is the essential supremum (assuming to existence) of $\phi\;|f^{p-2}f'|$ on the support of $f$. If $\alpha=\infty$ similar in \cite{LYZ}, the $J^{\rm w,\phi}_{\alpha,p}(f)$ is given by
\beq\label{Eq:2.1} \diy V(\phi\; f^{p}\big/p)-\diy\int\phi'\; f^{p}\big/p\; \rd x,\eeq
where $V(f)$ is the total variation (here assuming $f^{p}$ has bounded variation) and $\phi'$ denotes the derivative of WF $\phi$, $\diy\phi'(x)=\diy\frac{\rd}{\rd x}\phi(x)$. It can be easily seen that when $\phi\equiv 1$ the integral in (\ref{Eq:2.1}) vanishes. For more details, we once more address the reader \cite{F, S, Z, FU, Be1, Be2}.
\end{defn}

\begin{rem}
In special form $\varphi=f^k|f'|^m$, $k\in\bbR$, $m>1-\beta$ one obvious formula reads
\beq J^{\rm w,\phi}_{\alpha,p}(f)=\bigg(J_{\alpha',p'}(f)\bigg)^{\beta' p'},\eeq
where $\alpha'$ is the H\"{o}lder conjugate of $\beta'$ such that
$$\beta'=m+\beta, \quad p'=(k+p\beta+2m)\big/(m+\beta).$$
If we deal with the WF $\phi$ as a polynomial function of $f$: $\varphi=\diy\sum_{i=0}^n b_i\;f^i$ where $b_i,\;i=0\dots n$ are constant and $\varphi\geq0$. Then one has
\beq\label{Exa.WFI} J^{\rm w,\phi}_{\alpha,p}(f)=\diy \sum_{i=0}^n b_i\bigg(J_{\alpha,p_i}(f)\bigg)^{\beta p_i}, \quad p_i=p+\frac{i}{\beta}.\eeq
Note that $J_{.,.}(f)$ stands as before in (\ref{Def:FI}). By looking at (\ref{Exa.WFI}) it's not difficult to deduce the $(\alpha,p)$-th WFI is not necessary decreasing in $\alpha$ for given $p$.
\end{rem}

In what follows, we will also use notation $\bbE(X)$ for the expectation relative to RV $X$ with the PDF $f$. For arbitrary RV $Z$ and given WF $\phi$,  set
\beq\begin{array}{cl} \diy\widetilde{\Lambda}_{\phi,p}(Z)=\diy\bbE\bigg\{\phi\Big(-\Big\{\frac{(1-Z)}{(p-1)}\Big\}^{1/\alpha}\Big)
+\phi\Big(\Big\{\frac{(1-Z)}{(p-1)}\Big\}^{1/\alpha}\Big)\bigg\},\\
\\
\diy\overline{\Lambda}_{\phi,p}(Z)=\diy\bbE\bigg\{\phi\Big(-\Big\{\frac{(1-Z)}{Z\;(1-p)}\Big\}^{1/\alpha}\Big)
+\phi\Big(\Big\{\frac{(1-Z)}{Z\;(1-p)}\Big\}^{1/\alpha}\Big)\bigg\}.\end{array}\eeq
Assuming all expectations are finite, the notations below are proposed allowing us to shorten the formulas throughout the paper:
\beq\label{Eq:2.7} \Theta_{\alpha}(Z)=\bbE\Big\{\phi(-Z^{1/\alpha})+\phi(Z^{1/\alpha})\Big\},\qquad \Upsilon(Z)=\diy\bbE\Big\{\phi(-e^{-Z})+\phi(e^{-Z})\Big\}.\eeq

Next, by virtue of Definitions (\ref{Def:1.1}), (\ref{Def:1.2}) and (\ref{Def:2.1}), a list of weighted information measures for PDF $G$ in various cases of $\alpha$, $p$ is established:
\begin{itemize}
\item[$\bullet$] For $\alpha\in(0,\infty)$ and $p>1$, we start with an explicit form of $N^{\rm w}_{\phi,p}(G)$. Therefore after not complicated computations, one obtains
\beq N^{\rm w}_{\phi,p}(G)=\diy a_{\alpha,p}^{-1}\;2^{1/(p-1)}\;\Big(\frac{p\;\alpha}{p\;\alpha+p-1}\Big)^{1/(1-p)}\;\big\{\widetilde{\Lambda}_{\phi,p}(Z)\big\}^{1/(1-p)},\eeq
where $Z\sim {\rm Beta}\;\Big(1/\alpha,(2p-1)/(p-1)\Big)$. Also assume RV $Y$ has Beta distribution with parameters $p/(p-1)$, $(\alpha+1)/\alpha$. Then owing to (\ref{Def:sigma}), one can write
\beq \sigma_{\phi,\alpha}(G)=\Big\{\big(2\;(p\alpha+p-1)\big)^{-1}\;\widetilde{\Lambda}_{\phi,p}(Y)\Big\}^{1/\alpha}.\eeq
For a subset of $\alpha$ as $[1,\infty)$ and $p>1$, observe
\beq\label{eq:Fisher.1} J^{\rm w,\phi}_{\alpha,p}(G)=\big(2(\alpha p+p-1)\big)^{-1}\; a_{\alpha,p}^{\beta(p-1)}\;\alpha^{\beta}.\widetilde{\Lambda}_{\phi,p}(Y).\eeq
Note that here $\beta$ is the H\"{o}lder conjugate of $\alpha$. Consequently we have the following assertion involving $N^{\rm w}_{\phi,p}(G)$, $\sigma_{\phi,\alpha}(G)$ and $J^{\rm w,\phi}_{\alpha,p}(G)$:
\beq\label{Eq:2.11} \diy\big[N^{\rm w}_{\phi,p}(G)\big]^{1-p}=\diy p\;\sigma_{\phi,\alpha}(G)\;\big[J^{\rm w,\phi}_{\alpha,p}(G)\big]^{1/\beta}\;\diy\frac{\widetilde{\Lambda}_{\phi,p}(Z)}{\widetilde{\Lambda}_{\phi,p}(Y)}.\eeq
\item[$\bullet$] For $\alpha\in(0,\infty)$ and $p\in(1/(\alpha+1),1)$, consider two RVs $\overline{Z}$ and $\overline{Y}$ having Beta PDFs
with parameters $p(\alpha+1)-1)/\alpha(1-p)$, $1/\alpha$ and $(p(\alpha+1)-1)/\alpha(1-p)$, $(\alpha+1)/\alpha$ respectively. With analogue manner one has
\beq\label{Eq:2.12} N^{\rm w}_{\phi,p}(G)=\diy a_{\alpha,p}^{-1} 2^{1/(p-1)}\;\Big(\frac{p\alpha}{p\alpha+p-1}\Big)^{1/(1-p)}\;\Big\{\overline{\Lambda}_{\phi,p}(\overline{Z})\Big\}^{1/(1-p)},\eeq
and the expression (\ref{Def:sigma}) becomes
\beq \sigma_{\phi,\alpha}(G)=\Big\{\big(2\;(p\alpha+p-1)\big)^{-1}\;\overline{\Lambda}_{\phi,p}(\overline{Y})\Big\}^{1/\alpha}.\eeq
Substituting $\overline{\Lambda}_{\phi,p}(\overline{Y})$ in $\widetilde{\Lambda}_{\phi,p}({Y})$ in (\ref{eq:Fisher.1}) and taking into account the previous arguments, $J^{\rm w,\phi}_{\alpha,p}(G)$ is derived when $\alpha\in[1,\infty)$. Thus
\beq \diy\big[N^{\rm w}_{\phi,p}(G)\big]^{1-p}=\diy p\;\sigma_{\phi,\alpha}(G)\;\big[J^{\rm w,\phi}_{\alpha,p}(G)\big]^{1/\beta}\;\diy\frac{\overline{\Lambda}_{\phi,p}(\overline{Z})}{\overline{\Lambda}_{\phi,p}(\overline{Y})}.\eeq
\item[$\bullet$] Let $\alpha\in(0,\infty)$ and $p=1$. Suppose RVs $W$, $\overline{W}$ have Gamma distribution with same rate parameter 1 but different scale parameters $(\alpha+1)/\alpha$, $1/\alpha$. To specify the $p$-th WRE, one yields
\beq h^{\rm w}_{\phi}(G)=\diy\bbE_{G}[\varphi] \big(\log\;a_{\alpha,1}^{-1}\big)+\diy(2\;\alpha)^{-1}\Theta_\alpha(W).\eeq
Consequently, it can be seen
\beq \begin{array}{cl} N^{\rm w}_{\phi,1}(G)=\diy a_{\alpha,1}^{-1}\;\exp\Big\{\Theta_\alpha(W)\Big/\alpha\;\Theta_\alpha(\overline{W})\Big\},\\
\\
 \sigma_{\phi,\alpha}(G)=(2\;\alpha)^{-1/\alpha} \big\{\Theta_{\alpha}(W)\big\}^{1/\alpha}.\end{array}\eeq
Recalling (\ref{def.GWFI}) once again, one can write a representation for the $(\alpha,1)$-th WFI:
\beq J^{\rm w,\phi}_{\alpha,1}(G)=\diy 2^{-1}\;\alpha^{\beta-1}\Theta_{\alpha}(W).\eeq
Here $\Theta_{\alpha}$ stands as the formula in (\ref{Eq:2.7}). Also it can be checked
\beq\label{Eq:2.18} \diy 2\;\sigma_{\phi,\alpha}(G)\big[J^{\rm w,\phi}_{\alpha,1}(G)\big]^{1/\beta}=\Theta_\alpha(W).\eeq
\item[$\bullet$] Case $\alpha=0$ and $p>1$. In a modified setting Consider three RVs $X, \overline{X}, \widetilde{X}$ having Gamma distribution with shape parameters $(2p-1)/(p-1)$, $1/(p-1)$, $p/(p-1)$ respectively and rate parameter $1$. Then one gives
\beq\begin{array}{l} N^{\rm w}_{\phi,p}(G)=\diy a_{0,p}^{-1}\;\Big(\frac{p}{2(p-1)}\Big)^{1/(1-p)}\;\Big\{\Upsilon(X)\Big\}^{1/(1-p)},\\
\diy\sigma_{\phi,0}(G)=\diy\exp\Big\{-(p-1)\;\Upsilon(\overline{X})
\big/\Upsilon(\widetilde{X})\Big\}.\end{array}\eeq
\item[$\bullet$] For $\alpha=\infty$ and $p>0$, set $\psi(x)=\diy \int_0^{x}\phi(t)\;\rd t$. According to (\ref{WREP}) one has the respective formula
\beq N^{\rm w}_{\phi,p}(G)=\diy 2^{p/(p-1)}\;\big(\psi(1)-\psi(-1)\big)^{1/(1-p)}.\eeq
Finally (\ref{Def:sigma}) admits the representation $\sigma_{\phi,\infty}(G)= esssup\;\phi(x)$. Regarding to conclude this part, for derivative function $\phi'$, set $\overline{\psi}(x)=\diy\int_0^x \phi'(t)\;\rd t$, then
\beq J^{\rm w,\phi}_{\infty,p}(G)=(p\;2^p)^{-1}\big[\psi(1)-\psi(-1)\big]-2^{-1-p}\big[\overline{\psi}(1)-\overline{\psi}(-1)\big].\eeq
Consequently
\beq \big[N^{\rm w}_{\phi,p}(G)\big]^{1-p}=\diy p\;J^{\rm w,\phi}_{\infty,p}(G)-p\;2^{-1-p}\big[\overline{\psi}(1)-\overline{\psi}(-1)\big].\eeq
\end{itemize}

Eventually for $t>0$ define $G_t:\bbR\rightarrow [0,\infty)$ as the form
\beq\label{PDF:Gt} G_t(x)=\diy \frac{1}{t} G(\frac{x}{t}),\eeq
which later in Section 3 will be used.

\section{Main extended inequalities}
We start with an extension of the moment-entropy inequality which will be applied to obtain an extended form for the Cram\'{e}r-Rao inequality. Further, let us discuss a kind of general version of the Fisher information inequality reflecting the properties of WFIs. In essence, Theorems 3.1-3.3 are deployments of their counterparts from \cite{LYZ}. Throughout this section we use a number of properties established in Sections 1 and 2.
\begin{thm}\label{thm:2}
{\rm{(The extended moment-entropy inequality (MEI) cf. \cite{LYZ}, Theorem 2.)}}\;Consider $f$ is any PDF. For given WF $\phi$ and  $G$ represented in Section 2, set
\beq\label{def:phi*}\phi^*(x)=\phi\bigg(\diy\frac{\sigma_{\phi,\alpha}(f)}{\sigma_{\phi,\alpha}(G)}\; x\bigg).\eeq
Here $\sigma_{\phi,\alpha}(.)$ represents the $\alpha$-th generalized deviation. Consider $\alpha\in[0,\infty]$, $p>1/(1+\alpha)$  and the following assumptions hold:
\beq\label{assum:1} \bbE_f[\phi]\geq \bbE_G[\phi]\;\;\; \hbox{and together with}\;\;\; \bbE_f[\phi]\geq \bbE_G[\phi^*],\;\; \hbox{if}\;\;p=1.\eeq
Then
\beq\label{Theorem:2} \diy \frac{N^{\rm w}_{\phi,p}(f)}{\sigma_{\phi,\alpha}(f)}\leq \diy \frac{\bigg(N^{\rm w}_{\phi,p}(G)\bigg)^p\bigg(N^{\rm w}_{\phi^*,p}(G)\bigg)^{1-p}}{\sigma_{\phi,\alpha}(G)}.\eeq
With equality if and only if $f\equiv G$.
\end{thm}

{\bf Proof:} Following arguments in Theorem 2, cf. \cite{LYZ}, we provide the proof in different cases: first for simplicity set $a=a_{\alpha,p}$ and
\beq\label{def:t} t_{\phi}=\diy\frac{\sigma_{\phi,\alpha}(f)}{\sigma_{\phi,\alpha}(G)}.\eeq
{\bf Case 1:} $\alpha\in(0,\infty)$ and $p\neq 1$. Owing to (\ref{PDF:G}) and (\ref{PDF:Gt}) one can write
\beq \label{case:1}\begin{array}{l}
\diy \int_{\bbR} \phi \; G_{t_\phi}^{p-1} f\;\rd x\\
\qquad\diy \geq a^{p-1}t_\phi^{1-p}\diy \int_{\bbR} \phi(x) f(x)\;\rd  x+(1-p)a^{p-1}t_\phi^{1-p-\alpha}\int_{\bbR}\phi(x)|x|^\alpha f(x) \rd x\\
\qquad =\diy a^{p-1}t_\phi^{1-p}\big(\bbE_f[\phi]+(1-p)t_\phi^{-\alpha}\mu_{\phi,\alpha}(f)\big),
\end{array}\eeq
Going back to (\ref{assum:1}), the RHS of (\ref{case:1}) is greater and equal than
\beqq \diy a^{p-1}t_\phi^{1-p}\big(\bbE_G[\phi]+(1-p)\mu_{\phi,\alpha}(f)\big)=\diy t_\phi^{1-p}\diy \int_{\bbR}\phi\; G^p\;\rd x.\eeqq
Note that the equality holds if $p<1$ and equality occurs in (\ref{assum:1}).\\
\\
{\bf Case 2:} $\alpha=\infty$ and $p\neq 1$. Observe that when $\alpha=\infty$, $f$ vanishes outside of interval $[-t_\phi,t_\phi]$. So one derives
\beq\label{case:2} \begin{array}{ccl} \diy \int_{\bbR}\phi\; G_{t_\phi}^{p-1}\;f\; \rd x &=& a^{p-1}t_\phi^{1-p}\int_{-t_\phi}^{t_\phi} \phi f\;\rd x\\
&\geq& a^{p-1}t_\phi^{1-p} \bbE_G[\phi]=t_{\phi}^{1-p}\diy \int _{\bbR}\phi\; G^{p}\;\rd x.\end{array}\eeq
Here the inequality holds because of (\ref{assum:1}). The scaling identity also can be checked for $G_{t_\phi}$:
\beq \label{scal.Gt}\diy \int_{\bbR}\phi(x)G_{t_\phi}^p(x)\;\rd x=t_{\phi}^{1-p}\diy \int_{\bbR}\phi(t_\phi x) G^p(x)\;\rd x.\eeq
With regard to $\phi^*(x)=\phi(t_\phi \;x)$, recalling (\ref{case:1}), (\ref{case:2}) and (\ref{scal.Gt}), one yields
\beq\label{eq:case2} \begin{array}{ccl} 1 &\leq& \diy \bigg(N^{\rm w}_{\phi,p}(f,G_{t_\phi})\bigg)^p\\
&=&\bigg(\diy \int_{\bbR}\phi\;G_{t_\phi}^p\;\rd x\bigg)\bigg(\diy \int_{\bbR} \phi\; f^p\;\rd x\bigg)^{\frac{-1}{1-p}}\bigg(\diy \int_{\bbR}\phi\; G_{t_{\phi}}^{p-1} f\;\rd x\bigg)^{\frac{p}{1-p}}\\
&=& \diy t_\phi \bigg(\diy\int_{\bbR}\phi\; G^p\;\rd x\bigg)^{\frac{p}{1-p}}\bigg(\diy\int_{\bbR}\phi^*\;G^p\;\rd x\bigg). N_{\phi,p}^{-1}(f)\\
&\leq & \diy \diy\frac{\sigma_{\phi,\alpha}(f)}{\sigma_{\phi,\alpha}(G)}.\frac{\bigg(N^{\rm w}_{\phi,p}(G)\bigg)^p\bigg(N^{\rm w}_{\phi^*,p}(G)\bigg)^{1-p}}{N^{\rm w}_{\phi,p}(f)},
\end{array}\eeq
implying (\ref{Theorem:2}). \\
\\
{\bf Case 3:} $\alpha\in (0,\infty)$ and $p=1$. By virtue of (\ref{assum:1}) and the Gibbs inequality in \cite{SY} one has
\beqq 0\leq \diy \frac{D^{\rm w}_{\phi}(f\|G_{t_\phi})}{\bbE_f[\phi]}=-\diy \frac{h^{\rm w}_{\phi}(f)}{\bbE_f[\phi]}-\log \;a +\log \; t_{\phi}+t_{\phi}^{-\alpha}\diy\frac{\mu_{\phi,\alpha}(f)}{\bbE_f[\phi]}.\eeqq
Using (\ref{assum:1}) and (\ref{def:t}) we conclude the required result in this case by
\beqq 0\leq -\diy \frac{h^{\rm w}_{\phi}(f)}{\bbE_f[\phi]}+\diy \frac{h^{\rm w}_{\phi}(G)}{\bbE_G[\phi]}+\log\; \sigma_{\phi,\alpha}(f)-\log\; \sigma_{\phi,\alpha}(G).\eeqq
{\bf Case 4:} $\alpha=\infty$ and $p=1$. With similar analogue method in case 3, one obtains
\beqq \begin{array}{l}
0\leq \diy \frac{D^{\rm w}_{\phi}(f\|G_{t_\phi})}{\bbE_f[\phi]} = -\diy \frac{h^{\rm w}_{\phi}(f)}{\bbE_f[\phi]}-\log \;a +\log\; t_{\phi}\\
\quad= -\diy \frac{h^{\rm w}_{\phi}(f)}{\bbE_f[\phi]}+\diy \frac{h^{\rm w}_{\phi}(G)}{\bbE_G[\phi]} +\log\; \sigma_{\phi,\infty}(f)-\log\; \sigma_{\phi,\infty}(G).\end{array}\eeqq
{\bf Case 5:} $\alpha=0$ and $p>1$. Owing to (\ref{Def:sigma}):
\beq \diy\int_{\bbR}\phi\;G^{p} \rd x=-a^{p-1} \bbE_{G}(\phi)\log\;\sigma_{\phi,0}(G).\eeq
By virtue of (\ref{assum:1}), (\ref{Def:sigma}) and (\ref{def:t}), it turns out
\beq \begin{array}{l}\diy\int_{\bbR}\phi\; G_{t_\phi}^{p-1} f\;\rd x\\
\quad \geq \diy \bbE_f[\phi]\bigg(t_{\phi}^{1-p}a^{p-1}\log \;t_{\phi}-t_{\phi}^{1-p}a^{p-1}\log \;\sigma_{\phi,0}(f)\bigg)\\
\quad =\diy t_{\phi}^{1-p} \diy\int_{\bbR}\phi\; G^p\; \rd x\bigg(\diy \frac{\bbE_f(\phi)}{\bbE_G(\phi)}\bigg)\\
\quad \geq t_{\phi}^{1-p}\diy \int_{\bbR}\phi\; G^{p}\;\rd x.\end{array}\eeq
Taking into account (\ref{eq:case2}) the result is verified. According to the Gibbs inequality, the equality occurs iff $f\equiv G_{t_\phi}$, for some $t_{\phi}\in(0,\infty)$, where this is implied from $f\equiv G$.  $\quad$ $\blacksquare$

\vspace{0.5 cm}
In addition, an immediate application of Theorem \ref{thm:2} by choosing $p=\alpha=1$ in Corollary \ref{cor1} can be established as the following:
\begin{cor}\label{cor1}
Consider RV $X$ with the PDF $f$. Let $\phi(x)=|x|^c$, $c\in\bbR$. For $\alpha\in[0,\infty]$, set
\beq t_c(f,G):=\bigg(\diy \frac{\sigma_{c+\alpha}(f)}{\sigma_{c+\alpha}(G)}\bigg)^{\diy c(c+\alpha)/\alpha}.\eeq
Suppose that PDF $f$ obeys
\beq \begin{array}{l}\diy\bbE_f\big[|X|^c\big]\geq \bbE_G\big[|X|^c\big],\quad\hbox{and together with,}\\
\quad \diy\bbE_f\big[|X|^c\big]\geq \bbE_G\Big[t_c(f,G)\big|X\big|^c\Big],\;\;\hbox{if}\;\; p=1.\end{array}\eeq
Then for $p>1/(1+\alpha)$
\beq \diy \frac{\sigma_{\phi,\alpha}(f)^{c+1}}{N^{\rm w}_{\phi,p}(f)}\geq \frac{\sigma_{\phi,p}(G)^{c+1}}{N^{\rm w}_{\phi,p}(G)},\quad \hbox{equivalently}\quad \diy \frac{\sigma_{c+\alpha}(f)^{C_\alpha}}{N^{\rm w}_{\phi,p}(f)}\geq \frac{\sigma_{c+\alpha}(G)^{C_\alpha}}{N^{\rm w}_{\phi,p}(G)}.\eeq
Here $C_\alpha=(c+1)(c+\alpha)/\alpha$. Consequently following \cite{CT, LYZ}, the appropriate extremal distribution maximizes
\beqq N^{\rm w}_{\phi,p}(f)=\bigg(\diy\int_{\bbR}|x|^c f^p(x) \rd x\bigg)^{1/(1-p)}.\eeqq
with the same $c+\alpha$-th moment. On the other hand a direct assertion can be expressed as the following:
for any PDF $f:\bbR\mapsto\bbR$ where satisfies in suppositions
\beq\label{eq:3.13} \bbE_f[|X|^c]\geq c!,\quad \bbE_f[|X|^c]\geq \diy\frac{1}{c+1}\Big(\bbE_f[|X|^{c+1}]\Big)^c, \quad c\in\bbR.\eeq
One yields
\beq\label{eq:3.14}\diy\frac{(c+1)!\;N^{\rm w}_{|x|^c,1}(f)}{2\;e^{c+1}}\leq \diy\int_{\bbR}|x|^{c+1} f(x)\rd x.\eeq
The equality occurs when $f\equiv\diy \frac{1}{2} e^{-|x|}$. Note that for instance if $f\sim Exp(\lambda)$ for chosen ranges $\lambda\in(3,\infty)$, $c\in(-1,0)$,  both inequalities in (\ref{eq:3.13}) are fulfilled. This implies (\ref{eq:3.14}) as one of the variate possible bounds for WRP.
\end{cor}
\vspace{0.1 cm}
In this stage the author analyses another particular case $p=2,\alpha=1$ in Theorem \ref{thm:2}, in order to explore one more result below.
\begin{cor}\label{cor2} Define the quantity $m(c)$ by
$$\diy\frac{2^{2-c}}{(c+3)^{2-c}\;(c+2)^{1-c}}.$$
Under condition
\beqq \bbE_f[|X|^c]\geq \diy\frac{2}{(c+2)(c+1)},\quad c+2>0 \eeqq
One has
\beq m(c)\;\bigg(\diy\int_{\bbR}|x|^{c+1}f(x)\; \rd x\bigg)^{c-1}\leq \diy\int_{\bbR}|x|^c f^2(x)\;\rd x.\eeq
Equality holds if and only if $f\equiv \big(1-|x|\big)_+$.
\end{cor}
Next the reader is referred to the definitions of the $(2,2)$-th weighted FI and the $(2,2)$-th FI introduced in (\ref{Def:FI}). By Taking into account the $2$-th WRP and the $2$-th generalized deviation for given WF $\phi(x)=\big(G'(x)\big)^2$, we pass to the following assertion.
\begin{cor}
Consider the generalized $2$-Gaussian PDF, $G(x)=\diy\frac{3}{4}\big(1-x^2)$, $x\in(-1,1)$. Let $f:\bbR\mapsto \bbR$ be a PDF satisfied in
\beq \label{eq1:cor3}\sigma_2(f)\geq \diy\frac{2}{3}\Big(J_{2,2}(G)\Big)^2.\eeq
Define constant $w(G)$ by
\beqq \Big(J_{2,5/2}(G)\Big)^{10}\;\Big(J^{\rm w, x^2}_{2,2}(G)\Big)^{-3/2}.\eeqq
Then the inequality
\beq \bigg(\diy\int_{\bbR}x^2\;f^2(x)\;\rd x\bigg)^{-1}\leq 2\;w(G)\;\bigg(\diy\int_{\bbR} x^4\; f(x)\;\rd x\bigg)^{3/2},\eeq
holds true. The equality occurs when $f\equiv \diy\frac{3}{4}\big(1-x^2)_+$. Let us note that $\sigma_2(f)$ in (\ref{eq1:cor3}) is $2$-th deviation given by
\beqq \sigma_2(f)=\bigg(\diy\int_{\bbR} x^2\;f(x)\;\rd x\bigg)^{1/2}.\eeqq
\end{cor}

Recently, in \cite{LYZ}, it has been shown that among all PDFs, the unique distribution that minimizes $p$-th Renyi entropy with $(\alpha,p)$-th Fisher information is Gaussian. Regarding to the weighted version, employing the WF $\phi$, we establish an extended assertion associated the generalized Gaussian. The proof is given in Appendix.

Suppose RV $X$ has PDF $f$. let $(a,b)$, $a,b\in [-\infty,\infty]$ be the smallest interval containing the support of absolutely continuous PDF $f$. Define an increasing absolutely continuous function $s:(a,b)\mapsto (-k,k)$, for some $k\in(0,\infty]$ such that $\forall x\in(a,b)$
\beqq \diy\int_a^x f(t)\rd t=\diy\int_{-k}^{s(x)} G(t)\rd t.\eeqq
Here $G$ represents the generalized Gaussian density. Observe that RV $S:=s(X)$ has density $G$. Next consider a WF $x\in\bbR\mapsto \phi(x)\geq 0$. Given $p$, $\alpha$ and its H\"{o}lder conjugate $\beta$, introduce additional WFs:
\beq\label{def:rho1.2} \rho_1(x)=\big(\phi(x)\big)^{\alpha/(1-p)},\;\;\; \rho_2(x)=\big(\phi(x)\big)^{p\beta/(p-1)}.\eeq
Further, let $x\in\bbR\mapsto T(x)\in \bbS$ be an differentiable function. Denoting, as before, the derivation of function $\rho$ by $\rho'$, define
\beq\label{Def.RT} \rho_s(x)=\big(\widetilde{\phi}(x)\big/\phi^p(x)\big)^{1/(1-p)}, \quad \eta_{\phi,p}(s)=\diy\int_{\bbS} s(x) \rho'_s(x) f^p(x) \rd x,\eeq
where $\widetilde{\phi}(x)=\phi(s(x))$. Here
\beqq \rho'_s(x)=\rho_s(x)\;\left[\diy\frac{1}{1-p}\left(\frac{s'(x) \phi'(s(x))}{\phi(s(x))}\right)+\frac{p}{p-1}\left(\frac{\phi'(x)}{\phi(x)}\right)\right].\eeqq
Moreover, involving the $p$-th WRP of $G$ given in Section 2, set
\beq\label{Def:kappa} \kappa_{\phi,p}=\diy \eta_{\phi,p}(s)\big[N^{\rm w}_{\rho_1,p}(G)\big]^{p-1}.\eeq
One deduces that when $\phi\equiv 1$, $\eta_{\phi,p}(s)$ and consequently $\kappa_{\phi,p}(s)$ are removed.
\vspace{0.5 cm}
\begin{thm}\label{thm:3}
{\rm{(The extended Fisher information inequality (FII) cf. \cite{LYZ}, Theorem 3.)}}\;Assume $\alpha<\infty$, $p>1$. Also consider $1/\alpha+1/\beta=1$, two RVs $Z$, $Y$ invoking Beta distributions with the first shape parameter $1/\alpha$, $p/(p-1)$ and the second shape parameters $ (2p-1)/(p-1)$, $(\alpha+1)/\alpha$ respectively. Hence given continuous WF $\phi$
\beq\label{ExtendF:1} \Big[\diy \frac{N^{\rm w}_{\phi,p}(G)}{N^{\rm w}_{\rho_1,p}(G)}\Big]\;\Big[\frac{N^{\rm w}_{\rho_1,p}(G)}{N^{\rm w}_{\phi,p}(f)}\Big]^{p}\leq
\Big[\frac{J^{\rm w,\rho_2}_{\alpha,p}(f)}{J^{\rm w,\rho_1}_{\alpha,p}(G)}\Big]^{1/\beta}\;\Big[\frac{\widetilde{\Lambda}_{\rho_1,p}(Y)}{\widetilde{\Lambda}_{\rho_1,p}(Z)}\Big]-\kappa_{\phi,p}(s).\eeq
The $\kappa_{\phi,p}(s)$ refers to (\ref{Def:kappa}). If $\alpha<\infty$, $p\in(1/(1+\alpha),1)$ substitute $\overline{\Lambda}_{\rho_1,p}(Y)\big/\overline{\Lambda}_{\rho_1,p}(Z)$ in \\
$\widetilde{\Lambda}_{\rho_1,p}(Y)\big/\widetilde{\Lambda}_{\rho_1,p}(Z)$. Furthermore, if $p=1$ then
\beq\label{ExtendF:2} \begin{array}{l}\diy\bigg(N^{\rm w}_{\phi,1}(G)\;\bbE_G[\phi]\big/N^{\rm w}_{\widetilde{\phi},1}(f)\bigg)^{\diy\bbE_G[\phi]}\\
\qquad\quad\leq \diy2^{-1}\;\Big(J^{\rm w,\widetilde{\phi}}_{\alpha,1}(f)\big/J^{\rm w,{\phi}}_{\alpha,1}(G)\Big)^{1/\beta}\;\Theta_\alpha(W)-\bbE_f[S\;\widetilde{\phi}'].\end{array}\eeq
Providing $W\sim {\rm Gamma} ((\alpha+1)/\alpha,1)$, and reduced WF $\widetilde{\phi}(x)=\phi(s(x))$ where $\widetilde{\phi}'=\diy\frac{\rd}{\rd x}\;\widetilde{\phi}$. Finally in case $\alpha=\infty$:
\beq\label{ExtendF:3} \Big[\diy\frac{N^{\rm w}_{\phi,p}(G)}{N^{\rm w}_{\phi,p}(f)}\Big]^{p}\leq \diy\frac{J^{\rm w,\rho_s}_{\infty,p}(f)}{J^{\rm w,\phi}_{\infty,p}(G)}-\Delta_{\phi,p}.\eeq
Here
\beq\Delta_{\phi,p}=\Big(J^{\rm w,\phi}_{\infty,p}(G)\Big)^{-1}\;\Big\{p^{-1}\;\eta_{\phi,p}(s)-2^{-1-p}\big[\overline{\psi}(1)-\overline{\psi}(-1)\big]\Big\}.\eeq
where $\overline{\psi}$ is given by $\diy\int_0^x \phi'(t)\;\rd t$.
\end{thm}
\vskip .5 truecm
\begin{rem} Passing to $\phi\equiv 1$, the reduced Eqns. $\kappa_{\phi,p}(s)$, $\Delta_{\phi,p}$ and $\bbE_f\big[T\;\widetilde{\phi}'\big]$ are vanished. Thus as result all (\ref{ExtendF:1}), (\ref{ExtendF:2}), (\ref{ExtendF:3}) illustrate the same result as (22) in \cite {LYZ}.
\end{rem}
Next assertion follows directly from Theorem \ref{thm:3} in special case $\alpha=1$, $p=1$.
\begin{cor}
Let $X$ be a RV with density $f$. Moreover let $(a,b)$, $a,b\in[-\infty,\infty]$ be the support of $f$. For given map $s: (a,b)\mapsto (-k,k)$ for some $k\in(0,\infty]$, such that for each $x\in(a,b)$
\beq \diy\int_a^x f(t)\;\rd t=\diy\frac{1}{2}\diy\int_{-k}^{s(x)} e^{-|t|}\;\rd t. \eeq
For constant $c$, consider RV $S:s(X)$ and set
\beq A_s(f)=\diy\bbE_f\big[S^2\; S'\;|S|^{c-2}\big],\qquad B_s(f)=\diy\bbE_f\big[S\;S'e^{-c\;S}\big].\eeq
Then for $-1<c$ one has
\beq \diy\bigg(2\;c!\;\exp\{2^c\}\big/N^{\rm w}_{|s|^c,1}(f)\bigg)^{c!}\leq \diy c!\;2^c \sup|s|^c|(\log\;f)'|-c\; A_s(f).\eeq
And for $-\diy\frac{1}{2}<c<\diy \frac{1}{2}$ one gets
\beq \diy (1-4\;c^2)\bigg(2\exp\Big\{\diy\frac{1-c^2}{1-4\;c^2}\Big\}\big/(1-c^2)N^{\rm w}_{e^{-cs},1}(f)\bigg)^{1/(1-c^2)}\\
\quad\qquad \quad \leq  \diy \sup e^{-cs}|(\log f)'|+c\;(1-4\;c^2)\;B_s(f).\eeq
Here $s'$ denotes the derivative of $s$.
\end{cor}

An application of the extension version MEI and FII provides the following developed form of the Cram\'{e}r-Rao inequality:

\begin{thm}\label{thm:4}
{\rm{(The extended Cram\'{e}r-Rao inequality, cf. \cite{LYZ}, Theorem 5.)}}\;Assume the suppositions in Theorem \ref{thm:2}, \ref{thm:3}. For given WF $\phi$ the reduced WFs $\phi^*$, $\rho_1$ are considered as before, (\ref{def:phi*}), (\ref{def:rho1.2}). Moreover set
\beq \varpi_{\phi,p}(f,G)=\diy N^{\rm w}_{\phi^*,p}(G)\;N^{\rm w}_{\rho_1,p}(G)\big/N^{\rm w}_{\phi,p}(G)\;N^{\rm w}_{\phi,p}(f).\eeq
Then the LHS of inequalities (\ref{ExtendF:1}), (\ref{ExtendF:2}) and (\ref{ExtendF:1}) becomes
\beqq \diy \frac{\sigma_{\phi,\alpha}(G)}{\sigma_{\phi,\alpha}}\;\big(\varpi_{\phi,p}(F,G)\big)^{p-1},\;\;\;\hbox{if}\;\;\alpha<\infty,\; p>1,\eeqq
and if $p=1$,
\beqq \diy\Big(\sigma_{\phi,\alpha}(G)\;\bbE_{G}[\phi]\big/\sigma_{\phi,\alpha}(f)\Big)^{\diy\bbE_G[\phi]}.\eeqq
Further, in particular case $\alpha=\infty$, the LHS swaps by
\beqq \sigma_{\phi,\alpha}(G)/\sigma_{\phi,\alpha}(f).\eeqq
Note that we avoid to repeat the RHS of the corresponding inequalities.
\end{thm}
\vskip .5 truecm
{\emph{Acknowledgements --}}
SYS thanks the CAPES PNPD-UFSCAR Foundation
for the financial support in the year 2014-5. SYS thanks
the Federal University of Sao Carlos, Department of Statistics, for hospitality during the year 2014-5.

\vskip .5 truecm

{\LARGE{{\bf Appendix}}}\\

{\bf Lemma 4, cf. \cite{LYZ}:} Let $a,b\in[-\infty,\infty]$ and $f:(a,b)\mapsto \bbR$ be an absolutely continuous function such that
$$\lim\limits_{x\rightarrow a}f(x)=\lim\limits_{x\rightarrow b}f(x)=0.$$
Let map $g:(a,b)\mapsto\bbR$ be an increasing absolutely continuous function such that
$\lim\limits_{t\rightarrow b}g(t) >0,$ and the integral $\diy \int_a,b f'\;g\; \rd x$ is absolutely convergent. Then
\beqq \diy\int_a^b f\;g'\;\rd x=-\int_a^b f'\;g\; \rd x.\eeqq

{\bf Proof of Theorem 3.2:}
To implement the same steps as in the proof of Theorem \ref{thm:2}, cf \cite{LYZ} one shall offer the proof in three cases:\\
\\
{\bf Case1:} $p\neq 1$, $\alpha<\infty$. We begin this case via computing the $p$-th WRP for RV $S:=s(t)$:
\beq N^{\rm w}_{\phi,p}(G)=\left\{\begin{array}{ll}
                             \diy \bigg[\int_a^b\phi(s(x))\; f^{p}\; (s')^{1-p}\;\rd x \bigg]^{1\big/(1-p)}&  p\neq 1\\
                             \exp\Big\{\diy \frac{h^{\rm w}_\phi(g)}{\bbE_g[\phi]}\Big\} & p=1
                           \end{array}\right. \eeq
It can be easily seen that
\beqq  h^{\rm w}_\phi(G)=h^{\rm w}_{\widetilde{\phi}}(f)+\bbE_f\big[\widetilde{\phi}\;\log\;s'\big].\eeqq
Here $\widetilde{\phi}(z)=\phi(s(z))$ and $s'$ is derivative function of $s$. Now one has
\beq\label{Eq:5.2} \begin{array}{l} \bigg(N^{\rm w}_{\phi,p}(f)\bigg)^{-p} N^{\rm w}_{\phi,p}(G)\\
\quad =\bigg(\diy \int_a^b \phi\; f^p\;\rd x\bigg)^{-p\big/(1-p)}\;\bigg(\diy \int_a^b \phi(s(x))\;f^p\;(s')^{1-p}\;\rd x\bigg)^{1\big/(1-p)}\\
\qquad \leq \diy \int_a^b \rho_s(x)\; s'\;f^p\;\rd x.\end{array}\eeq
The inequality comes from the H\"{o}lder inequality with $\rho_s$ in (\ref{Def.RT}). Recall Lemma 4 cf. \cite{LYZ} and the notation in (\ref{Def.RT}), then the RHS of (\ref{Eq:5.2}) takes the form
\beqq -p\;\diy\int_a^b s(x) \rho_s(x)f^{p-1}(x) f'(x)\;\rd x -\eta_{\phi,p}(s).\eeqq
At this stage let us focus on the above integral:
\beqq \begin{array}{l}
-p\;\diy\int_a^b s(x) \rho_s(x)f^{p-1}(x) f'(x)\;\rd x \\
\quad\leq p\;\bigg(\diy \int_a^b (\widetilde{\phi}(x))^{\alpha/(1-p)}|s(x)|^\alpha f(x) \rd x\bigg)^{1/\alpha}\\
\qquad \; \bigg(\diy\int_a^b \phi(x)^{p\beta/(p-1)}|f(x)^{p-1-1/\alpha}\; f'(x)|^{\beta} \rd x\bigg)^{\1/\beta}\\
\quad\leq p.\sigma_{\rho_1,\alpha}(G)\;\big[J^{\rm w, \rho_2}_{\alpha,p}(f)\big]^{1/\beta}.\end{array}\eeqq
This leads
\beq \bigg(N^{\rm w}_{\phi,p}(f)\bigg)^{-p} N^{\rm w}_{\phi,p}(G)\leq \diy p\;\sigma_{\rho_1,\alpha}(G)\;\big[J^{\rm w, \rho_2}_{\alpha,p}(f)\big]^{1/\beta}-\eta_{\phi,p}(s).\eeq
Eventually using expressions (\ref{Eq:2.11}) and (\ref{Eq:2.12}) gives the required results. \\
\\
{\bf Case2:} $\alpha<\infty$, $p=1$. To deduce (\ref{ExtendF:2}) we use Jensen's inequality. One yields
\beqq\begin{array}{l}\hwphi(G)=h^{\rm w}_{\widetilde{\phi}}(f)+\diy \int_a^b \widetilde{\phi}(x) f(x) \log s'(x) \rd x\\
\quad \leq h^{\rm w}_{\widetilde{\phi}}(f)-\bbE_f[\widetilde{\phi}]\log\;\bbE_f[\widetilde{\phi}]+\bbE_f[\widetilde{\phi}]\log\diy\int_a^b \widetilde{\phi}(x) f(x)s'(x) \rd x.\end{array}\eeqq
By virtue of Lemma 4 in \cite{LYZ} and H\"{o}lder inequality once again, one obtains
\beqq\begin{array}{l} \hwphi(G)\\
\quad\diy\leq h^{\rm w}_{\widetilde{\phi}}(f)-\bbE_f[\widetilde{\phi}]\log\;\bbE_f[\widetilde{\phi}]+\log\Big\{\sigma_{\phi,\alpha}(G)\Big[J^{\rm w,\widetilde{\phi}}_{\alpha,1}(f)\Big]^{1/\beta}-\diy\bbE\big[S\;(\widetilde{\phi})'\big]\Big\}.\end{array}\eeqq
where $\beta$ as in entire of the paper is the H\"{o}lder conjugate of $\alpha$. Therefore applying (\ref{Eq:2.18}), (\ref{Eq:1.5}) and $\bbE_f[\widetilde{\phi}]=\bbE_G[\phi]$, completes the proof in this case.\\
\\
{\bf Case3:} $\alpha=\infty$. Setting $s(x)=-k$ for all $x\in (-\infty, a]$ and $s(x)=k$ for all $x\in [b,\infty)$, let us go back to (\ref{Eq:5.2}) which gives
\beqq \begin{array}{cl} \Big(N^{\rm w}_{\phi,p}(f)\Big)^{-p} N^{\rm w}_{\phi,p}(G) \leq \diy -\int_a^b s(x) \rho_s'(x) f^p(x)\rd x-\diy \int_a^b s(x) \rho_s(x) (f^p(x))' \rd x\\
\leq k\diy \int_{\bbR} \rho_s(x)|(f^p(x))'|\rd x-\eta_{\phi,p}(s)\\
=\diy p\;J^{\rm w,\rho_s}_{\infty,p}(f)-\eta_{\phi,p}(s).\end{array}\eeqq
Therefore the claimed bound is provided. $\qquad$ $\blacksquare$

\bibliographystyle{plain}

\begin{thebibliography}{10}
\bibitem{BG} M. Belis and S. Guiasu.  A Quantitative and qualitative measure of information in cybernetic systems. {\it IEEE Trans. on Inf. Theory},  \textbf{14} (1968), 593--594.

\bibitem{Be1} J. F. Bercher. On generalized Cram\'{e}r-Rao inequalities, generalized Fisher information and characterizations of generalized q-Gaussian distributions. {\it Journal of Physics A: Mathematical and Theoretical}, \textbf{Vol. 45} (2012), No. 25, 255--303.


\bibitem{Be2} J. F. Bercher. On a $(\beta, q)$-generalized Fisher information and inequalities involving q-Gaussian distributions. {\it J. Math. Phys.}, \textbf{Vol. 53}, (2012), no. 6, 063--303.



\bibitem{C} A. Clim. Weighted entropy with application. {\it Analele Universit\v{a}\c{t}ii Bucure\c{s}ti, Matematic\v{a}}, Anul \textbf{LVII} (2008), 223-231.

\bibitem{CHV} J. A. Costa, A. O. Hero and C. Vignat. A characterization of the multivariate distributions maximizing renyi entropy. {\it In proceedings of 2002 IEEE International Sumposium on Information Theory}, (2002), page 263.


\bibitem{CT} T. Cover and J. Thomas. {\it Elements of Information Theory.} New York: Wiley, 2006.

\bibitem{Cr} H. Cram\'{e}r. {Mathematical methods of statistics.} Princton Landmarks in Mathematics. Princton, N. J: Princton University Press, 1999, reprint of the 1946 original.




\bibitem{G} S. Guiasu. Weighted entropy. {\it Report on Math. Physics}, \textbf{2} (1971), 165--179.

\bibitem{DD} M. Del Pino and J. Dolbeault. Best constants for Gagliardo-Nirenberg inequalities and applications to nonlinear diffusions. {\it J. Math. Pures Appl.}, \textbf{Vol. 81} (2002), no. 9, 847--875.


\bibitem{F} R. A. Fisher. Theory of statistical estimation. {\it Philos. Trans. Roy. Soc. London Ser A}, \textbf{Vol. 222} (1930), 309--368.




\bibitem{FU} S. Furuichi. On generalized Fisher informations and Crem\'{e}r-Rao type inequalities. {\it Journal of Physics: Conference Series 2}, \textbf{201} (2010), 012--016.




\bibitem{KS2} M. Kelbert and Y. Suhov. {\it Information Theory and Coding by Example.} Cambridge: Cambridge University Press, 2013.





\bibitem{LYZ} E. Lutwak, D. Yang and G. Zhang. Cram\'{e}r-Rao and moment-entropy ineqialities for Renyi entropy and generalized Fisher information. {\it IEEE Transaction on Information Theory}, \textbf{Vol 51} (2005), no. 2, 473--478.


\bibitem{LYZ1} E. Lutwak, D. Yang and G. Zhang. Moment-entropy inequalities. {\it Annals of Probability}, \textbf{vol 32} (2004), 757--774.



\bibitem{Oh} J. R. Ohm. {\it Multimedia Comunication Technology.} Springer-Verlag Berlin Heidelberg, 2004.

\bibitem{R} A. Renyi. On measures of entropy and information. {\it In Proc. Fourth Berkeley Symp. Math. Stat. Prob.}, {\bf Vol 1} (1960), page 547. Berkeley, (1961). University of California Press.

\bibitem{ShMM} B. D. Sharma, J. Mitter and M. Mohan. On measure of  `useful` information. {\it Inform. Control} \textbf{39} (1978), 323--336.

\bibitem{SiB} R. P. Singh and J. D. Bhardwaj. On parametric weighted information improvement. {\it Inf. Sci.} \textbf{59} (1992), 149--163.


\bibitem{S} A. Stam. Some inequalities satisfied by the quantities of information of Fisher and Shannon. {\it Inform. Contr.}, {\bf Vol 2} (1959), 101--102.

\bibitem{SY} Y. Suhov, I. Stuhl, S. Yasaei Sekeh and M. Kelbert. Basic inequalities for weighted entropies. arXiv 1510.02184.

\bibitem{SuYS} Y. Suhov, S. Yasaei Sekeh. An extension of the Ky Fan inequality.
{\it arXiv:1504.01166}

\bibitem{SuYSSt} Y. Suhov, S. Yasaei Sekeh and  I. Stuhl. Weighted Gaussian entropy and determinant inequalities entropy. {\it arXiv:1502.02188}


\bibitem{SYK} Y. Suhov, S. Yasaei Sekeh and M. Kelbert. Entropy-power inequality for weighted entropy. arXiv: 1502.02188.

\bibitem{SuStKel} Y. Suhov, I. Stuhl, M. Kelbert. Weight functions and log-optimal investment portfolios. {\it arXiv:1505.01437}



\bibitem{Z} R. Zamir. A proof of the Fisher infromation inequality via a data processing argument. {\it IEEE Transaction on Information Theory}, \textbf{44}, No. 3 (1998), 1246--1250.

\end{thebibliography}

 
\vspace{0.5cm}
\noindent Salimeh Yasaei Sekeh\\
is with the Statistics Department, DEs,\\
of Federal\;University\;of\;S$\tilde{\rm a}$o\;Carlos (UFSCar),\\
S$\tilde{\rm a}$o\;Paulo, Brazil\\
E-mail: sa$_{-}$yasaei@yahoo.com
\vskip 15pt

\end{document}